\documentclass[11pt]{article}
\usepackage{Blois,epsfig}

\def\bea{\begin{eqnarray}}
\def\eea{\end{eqnarray}}

\newcommand{\BQ}{\begin{equation}}
\newcommand{\EQ}{\end{equation}}
\newcommand{\BQA}{\begin{eqnarray}}
\newcommand{\EQA}{\end{eqnarray}}
\newcommand{\be}{\begin{eqnarray}}
\newcommand{\ee}{\end{eqnarray}}
\newcommand{\NN}{\nonumber \\}
\newcommand{\del}{\partial}
\newcommand{\tr}{{\rm tr}}

\newcommand{\ket}[1]{\left.\left\vert #1 \right. \right\rangle}

\newcommand{\x}{{\bf x}}
\newcommand{\y}{{\bf y}}

\newcommand{\z}{{\bf z}}
\newcommand{\w}{{\bf w}}

\def\simge{\mathrel{%
   \rlap{\raise 0.511ex \hbox{$>$}}{\lower 0.511ex \hbox{$\sim$}}}}
\def\simle{\mathrel{
   \rlap{\raise 0.511ex \hbox{$<$}}{\lower 0.511ex \hbox{$\sim$}}}}
\def\bigs{\mathrel{
   \rlap{\raise 0.531ex \hbox{$>$}}{\lower 0.531ex \hbox{$<$}}}}

\begin{document}
\vspace*{1.5cm}
\title{PERTURBATIVE ODDERON IN THE COLOR GLASS CONDENSATE}
\author{\vspace*{-4mm}K. ITAKURA}
\address{Servide de Physique Th\'eorique, 
CEA/Saclay, 91191 Gif-sur-Yvette Cedex, France \footnote{Address 
after June 1st 2005: 
Institute of Particle and Nuclear Studies, High Energy 
Accelerator Research Organization (KEK), Ooho 1-1, Tsukuba, 
Ibaraki, 305-0801, Japan. 
E-mail: {\tt kazunori.itakura@kek.jp}}}
\maketitle\abstracts{
\vspace*{-7mm}We discuss the ``odderon" exchange at high energy
within the framework of the Color Glass Condensate (CGC).
}
\vspace*{-12mm}
\setcounter{equation}{0}
\section{General strategies in the Color Glass Condensate}
\vspace{-2mm}
In this talk, we provide the description of $C$-odd odderon exchange 
within the framework of the Color Glass Condensate (CGC).\cite{Odderon}
The theoretical framework for the CGC 
is intended to describe hadronic scatterings at 
very high energies.\cite{CGCreview}  
Below, we briefly explain its general strategies by using the 
scattering of a simple projectile in a reference frame where most 
of the total momentum of the scattering is carried by the target 
($\simeq$ the infinite momentum frame of the target). The same 
strategies are applicable
for the odderon exchange as we will discuss later.

\noindent{\bf \ \ 1.} When the scattering energy is 
large enough, one can treat the target as the CGC which is a {\it dense 
gluonic state} on two dimensional transverse plane.
The CGC provides a strong gauge field $\alpha^a(x)=A_a^+(x)$ which is random 
due to random motion of the valence partons. The randomness 
of the gauge field is governed by the weight function $W_\tau [\alpha]$, whose 
energy (or rapidity $\tau$) dependence is determined by the so-called
JIMWLK equation. 

\noindent{\bf \ \ 2.} One has to determine the {\it scattering operator}
which is associated with the specific scattering process. 
Since the scattering energy is high,
one can use the eikonal approximation where the 
transverse positions of incoming partons do not change 
after the scattering. Thus, 
one can easily obtain the $S$-matrix for a fixed configuration 
of the gauge field $\alpha$ by computing the matrix element 
$\langle {\rm out}; \alpha, \x,\y,\z,\cdots | 
{\rm in}; \alpha, \x,\y,\z,\cdots \rangle $
where $\x,\y,\z, \cdots$ stand for the transverse positions of the partons
in the projectile. The outgoing parton differs from the incoming parton
up to the gauge factor which is picked up while traversing the external 
gauge field. To obtain the final result, one further has to take an 
average of this $S$-matrix over the possible configurations of the gauge 
field with the weight function $W_\tau[\alpha]$.

\noindent {\bf \ \ 3.} Once the relevant operator is specified, 
one is able to 
compute the evolution equation for it from the JIMWLK equation. 
This evolution equation is a first order differential equation 
with respect to the rapidity $\tau$. In particular, if the operator 
is gauge invariant, the JIMWLK equation can be made simple so that
infrared finiteness becomes manifest. The use of this simplified 
version of the JIMWLK equation greatly reduces the manipulations 
to obtain the final result of the evolution equation.
If one solves the evolution equation, one should be able to 
know how the 
$S$-matrix changes under the increment of scattering energy.

The simplest physical example is the dipole-CGC scattering where 
the incoming and outgoing states are $q\bar q$ dipoles: 
$
\ket{{\rm in};\alpha}\sim \bar\psi_i^{\rm in}(\x)\psi_i^{\rm in}(\y)\ket{0},
\ \ket{{\rm out};\alpha}\sim \bar\psi_i^{\rm
out}(\x)\psi_i^{\rm out}(\y)\ket{0}. $
The relation between $\psi^{in}$ and $ \psi^{out}$  is given by 
$ \psi^{out}_i= (V^\dag_{\x})_{ij} \psi^{in}_j$ with 
$V^\dag_\x$ being the Wilson
line (in the fundamental representation) along the trajectory of the 
quark:\vspace{-2mm}
\BQ\label{Wilson}
\vspace{-2mm}
V^\dag_\x = {\rm P}\, {\rm exp} 
\left\{ig \int dx^- \alpha^a (x^-,\x) t^a\right\}.
\EQ 
 The physical $S$--matrix is obtained after
averaging over the random external field: \vspace{-2mm}
\BQ\label{Sdipole}\vspace{-2mm}
S_\tau(\x,\y)= \int {\cal D}\alpha\,  W_\tau[\alpha]\,
\langle
{\rm out};\alpha \vert {\rm in};\alpha \rangle =
\frac{1}{N_c}\langle \tr (V^\dag_{\x} V_{\y})
 \rangle_\tau. 
\EQ
This implies that the relevant operator for the dipole-CGC scattering
is given by $\frac{1}{N_c} \tr (V^\dag_{\x} V_{\y})$. 
The evolution equation for this operator is obtained by the JIMWLK equation,
and the result is called the Balitsky equation.

\vspace{-2mm}
\setcounter{equation}{0}
\section{Odderon operators in the CGC}
\vspace{-2mm}
We follow the general strategies mentioned above to describe $C$-odd "odderon"
exchange at high energies. This can be done by adding the information of the 
$C$ parities to the incoming and outgoing states. Below we discuss two
simple projectiles: a color dipole and 3-quark state.

\vspace{1mm}
\noindent{\it 2.1 The dipole-CGC scattering}\\
To single out $C$-even
(`pomeron') or $C$-odd (`odderon') exchanges, 
we take the $C$-odd and $C$-even dipole states as the incoming 
 and outgoing states, respectively.
Then the relevant operator for $C$-odd exchanges in the
dipole-CGC scattering is extracted and is given by \vspace{-2mm}
 \BQ\vspace{-2mm}
O(\x,\y)
\equiv \frac{1}{2iN_c}\tr
(V_{\x}^{\dagger}V_{\y}-V_{\y}^{\dagger}V_{\x})\,=\,-\, O(\y,\x).
\label{dipole_odderon}
 \EQ
Note that the $C$-odd contribution is the imaginary part
of the $S$-matrix element: 
$
\langle O(\x,\y)
\rangle_\tau =\Im {\rm m\,} S_{\tau}(\x,\y).
$
The corresponding $C$-{even}, pomeron exchange,
amplitude, $N(\x,\y)$, is identified with the
real part of the $S$-matrix: 
$
N(\x,\y)\equiv 1- \frac{1}{2N_c} \tr
(V_{\x}^{\dagger}V_{\y}+V_{\y}^{\dagger}V_{\x}).
$

 From perturbative QCD, we expect that the lowest order contribution 
to the odderon exchange is of the form 
$d^{abc}A_\mu^a(\x)A_\nu^b(\y)A_\rho^c(\z)$
with $d^{abc}=2 \tr(\{t^a,t^b\}t^c)$ being a totally symmetric 
tensor.\cite{Ewerz} 
The same structure indeed 
emerges from the CGC operator (\ref{dipole_odderon}) 
 in the weak-field limit. 
By expanding the Wilson lines (\ref{Wilson}) 
up to {\it cubic} order with respect to the field 
$\alpha$, 
one finds the expected structure 
($\alpha_{\x}^a=\int dx^- \alpha^a(x^-,\x)$)\vspace{-2mm}
\BQ\vspace{-2mm}
O(\x,\y) \simeq \frac{-g^3}{24N_c}
  d^{abc}\left\{ 3(\alpha_{\x}^a\alpha_{\y}^b \alpha_{\y}^c-
\alpha_{\x}^a\alpha_{\x}^b \alpha_{\y}^c) +(\alpha_{\x}^a \alpha_{\x}^b
\alpha_{\x}^c-\alpha_{\y}^a \alpha_{\y}^b \alpha_{\y}^c)\right\}.
 \label{odd} \EQ
Note that this combination of trilinear field operators
is gauge invariant by construction.

\vspace{1mm}
\noindent{\it 2.2 The 3-quark--CGC scattering}\\
\noindent
We now turn to the 3-quark--CGC scattering at high energies.
The 3-quark colorless  state may be given by the 
"baryonic" operator $\epsilon^{ijk}
\psi^i(\x)\psi^j(\y)\psi^k(\z)$, where $\epsilon^{ijk}$ is the complete
antisymmetric symbol ($i,j,k$=1,2,3). 
By using the same eikonal approximation as for the
dipole-CGC scattering, one obtains 
the "3-quark odderon operator" \vspace{-2mm}
\BQ\vspace{-2mm}
O(\x,\y,\z)=\frac{1}{3!2i}\left(\epsilon^{ijk}\epsilon^{lmn}
V^{\dagger}_{il}(\x)V^{\dagger}_{jm}(\y)V^{\dagger}_{kn}(\z)- {\rm
c.c.}\right).\label{3q_odderon}
 \EQ
This somewhat unfamiliar operator can be made considerably 
simple if one multiplies the identity 
$
\epsilon^{ijk}\epsilon^{lmn}
V_{il}(\w)V_{jm}(\w)V_{kn}(\w)=3!\det V(\w)=3!
$, 
and then chooses the arbitrary coordinate $\w$ to be one of the 
quark coordinates, say $\w=\z$. Namely, one can equivalently rewrite 
(\ref{3q_odderon}) as \vspace{-2mm}
 \BQ\vspace{-2mm}
O(\x,\y,\z) \,=\, \frac{1}{3!2i}\Bigl [ \tr
(V_{\x}^\dag V_{\z})\tr (V_{\y}^\dag V_{\z})
  -\tr (V_{\x}^\dag V_{\z} V_{\y}^\dag V_{\z})- {\rm c.c.}\Bigr].\label{3q_odderon2}
 \EQ
Furthermore, when two of the coordinates are the same, the 3-quark
odderon operator reduces to the dipole odderon operator,
Eq.~(\ref{dipole_odderon}). 
 \BQ
 O(\x,\z,\z)\,=\,O(\x,\z)\,=\,-O(\x,\x,\z). \qquad (N_c=3) \label{avo} 
\EQ
This is physically reasonable because the diquark state is equivalent
to an antiquark as far as color degrees of freedom are concerned.

In the weak-field approximation, 
one finds again a gauge invariant linear combination of trilinear 
field operators with the $d$-symbol:\vspace{-2mm}
\BQ\label{sy}\vspace{-2mm}
O(\x,\y,\z)\simeq
\frac{g^3}{144}d^{abc}
\left\{\!
(\alpha_{\x}^a-\alpha_{\z}^a)\! +\! (\alpha_{\y}^a-\alpha_{\z}^a)\!\right\}\!
\left\{\!
(\alpha_{\y}^b-\alpha_{\x}^b)\! +\! (\alpha_{\z}^b-\alpha_{\x}^b)\!\right\}\!
\left\{\!
(\alpha_{\z}^c-\alpha_{\y}^c)\! +\! (\alpha_{\x}^c-\alpha_{\y}^c)\!\right\}.
\EQ

\vspace{-2mm}

\setcounter{equation}{0}
\section{Odderon evolution}
\vspace{-2mm}
Once we know the relevant operators for the $C$-odd scattering amplitudes, 
we can apply the JIMWLK equation 
 to the operators to derive the evolution equations for them. 

\vspace{-2mm}
\subsection{The dipole--CGC scattering}
\noindent
For the dipole-CGC scattering, the evolution equations obeyed by
the average amplitudes $\langle N(\x,\y) \rangle_\tau$ and $\langle
O(\x,\y)\rangle_\tau$ can be easily 
derived from the Balitsky equation because the 
operators $N(\x,\y)$ and $O(\x,\y)$ are, respectively, the real part
and the imaginary part of the dipole-CGC scattering operator 
$(1/N_c)\tr(V_{\x}^{\dagger}V_{\y})$ which satisfies the Balitsky equation. 
Therefore, the respective equations
can be simply obtained by separating the real part and the imaginary
part in the Balitsky equation. The result is ($O_{\x\y}=O(\x,\y),\, 
N_{\x\y}=N(\x,\y)$)\vspace{-2mm}
 \BQA 
\frac{\partial}{\partial \tau} \langle O_{\x\y}\rangle_\tau\!\!
 &=&\!\!\frac{\bar\alpha_s}{2\pi}\int
 d^2\z\  {\cal M}_{\x\y\z}
\, \Big\langle O_{\x\z}+ O_{\z\y}- O_{\x\y} 
-O_{\x\z}N_{\z\y}-
N_{\x\z} O_{\z\y}\Big\rangle_\tau,\ \ \label{11} \\
\frac{\del}{\del \tau}\langle N_{\x\y} \rangle_\tau
\!\!&=&\!\!\frac{\bar\alpha_s}{2\pi}\int d^2\z\,
{\cal M}_{\x\y\z}
\, \Big\langle
N_{\x\z}+ N_{\z\y}- N_{\x\y}
-  N_{\x\z}N_{\z\y} +
 O_{\x\z} O_{\z\y} \Big\rangle_\tau,\label{22} 
\EQA
where we have defined the dipole kernel
$
{\cal M}_{\x\y\z}={(\x-\y)^2}/{(\x-\z)^2(\z-\y)^2} \, .
$
As is the case with the Balitsky equations, the equations
above do not close by themselves.
 In the weak-field limit, both of the evolution equations 
reduce to the (linear) BFKL equation. However, the BFKL equation 
for the odderon exchange must be solved with the antisymmetric condition 
(\ref{dipole_odderon}). 
Therefore, even if the evolution equations are the same, 
the respective solutions behave differently. In particular, 
it is known that the highest intercept of the $C$-odd BFKL solution
 is given by 1 which is smaller than the (hard) 
pomeron intercept.\cite{kov}


In the mean-field approximation, Eqs.~(\ref{11})--(\ref{22}) reduce
to a closed system of coupled, non-linear, equations for $\langle N
\rangle_\tau$ and $\langle O \rangle_\tau$ (which was named 
as WHIMIKS equation in Ref.~\cite{Motyka})\vspace{-2mm}
 \BQA 
\frac{\partial}{\partial \tau}
\langle O_{\x\y}\rangle_\tau \! \!\!&=&\!\!\!
 \frac{\bar\alpha_s}{2\pi}\int d^2\z\ 
{\cal M}_{\x\y\z} \label{evolution_odd_fact}
 \Bigl [ \langle O_{\x\z}\rangle_\tau + \langle O_{\z\y}
\rangle_\tau - \langle O_{\x\y}\rangle_\tau 
- \langle
O_{\x\z}\rangle_\tau \langle N_{\z\y}\rangle_\tau - \langle
N_{\x\z}\rangle_\tau \langle O_{\z\y}\rangle_\tau\Bigr ], \ \nonumber\\
 \frac{\del}{\del \tau}\langle
N_{\x\y} \rangle_\tau\!\!\!  &=&\!\!\!\frac{\bar\alpha_s}{2\pi}\int d^2\z
\ {\cal M}_{\x\y\z}\label{evolution_even_fact}
\Bigl[ \langle  N_{\x\z}\rangle_\tau + \langle
N_{\z\y}\rangle_\tau - \langle N_{\x\y}\rangle_\tau 
-\langle N_{\x\z} \rangle_\tau \langle N_{\z\y} \rangle_\tau +\, \langle
O_{\x\z}\rangle_\tau \langle O_{\z\y}\rangle_\tau\Bigr]. \nonumber
 \EQA
The first of these equations has been already proposed in
Ref.~\cite{kov}, as a plausible non-linear generalization of
the BFKL equation in the $C$-odd channel.
The second equation is
the Balitsky-Kovchegov equation supplemented by a new term describing
the merging of two odderons. 

 One of the significant consequences of the nonlinear effects 
in the factorized evolution equation (the 2nd equation)
is that the odderon amplitude $\langle O\rangle_\tau$ will 
decay into zero with increasing energy.
This is easily seen by 
noting that when the pomeron amplitude $\langle N\rangle_\tau$ 
is close to 1 (deeply in saturation regime), the nonlinear terms in 
Eq.~(\ref{evolution_odd_fact}) cancel the first two terms on the r.h.s. 
and the resulting equation for $\langle O\rangle_\tau$ simply implies 
decrease of the solution. 
Therefore, as one goes to higher energies, the 
odderon contribution becomes less and less important.

\vspace{-2mm}

\subsection{The 3-quark--CGC scattering}
A straightforward application of the JIMWLK equation 
 to the relevant operator (\ref{3q_odderon}) 
leads to the desired evolution equation. 
However, the resulting equation is highly complicated and it seems 
difficult to extract physical information from the result. 
Instead, in this talk, we rather show the evolution 
equation for the weak-field version  (\ref{sy}) of the 
operator. This is indeed sufficient for the comparison of our result 
with the conventional description of the odderon exchange, 
the linear BKP equation.\cite{bartels,pra}
After a straightforward but lengthy calculation, the following 
linear {\it closed} 
evolution equation for $\langle O_{\x\y\z}\rangle_\tau \equiv 
\langle O(\x,\y,\z)\rangle_\tau$ is obtained
 \BQA
\frac{\partial}{\partial \tau} \langle O_{\x\y\z}\rangle_\tau
&=&\frac{3\alpha_s}{4\pi^2} \int d^2\w\ {\cal M}_{\x\y\w}\, 
\Bigl( \langle
O_{\x\w\z}\rangle_\tau + \langle O_{\w\y\z}\rangle_\tau -\langle
O_{\x\y\z} \rangle_\tau\NN && \qquad\quad - \langle
O_{\w\w\z}\rangle_\tau -\langle O_{\x\x\w}\rangle_\tau-\langle
O_{\y\y\w}\rangle_\tau-\langle O_{\x\y\w}\rangle_\tau \Bigr)\NN &&+\,
\Big\{\, {\rm 2\ cyclic\ permutations}\, \Big\}. \label{prot}
 \EQA
Let us compare this result with the BKP equation. At first grance, 
our result (\ref{prot}) does not look equivalent to the BKP equation. 
In fact, within our framework, the BKP 
equation rather appears as the evolution equation for the 
3-point Green's function defined by 
$
f_\tau(\x,\y,\z)\equiv d^{abc}\langle \alpha^a_{\x}\alpha_{\y}^b\alpha_{\z}^c\rangle_\tau.
$
Indeed, the evolution equation for this Green's function reads
 \be
\frac{\partial}{\partial \tau} f_\tau(\x,\y,\z)\!\!\!\! &=&\!\!\!\!
  \frac{\bar{\alpha_s}}{4\pi} \int d^2\w {\cal M}_{\x\y\w} 
\Big(
f_\tau(\x,\w,\z)+f_\tau(\w,\y,\z)
-f_\tau(\x,\y,\z)
 -f_\tau(\w,\w,\z) \Big)\NN
&& +\,\,\Big\{\, {\rm 2\  cyclic\  permutations}\, \Big\}\,.
 \label{eqf3} \ee
Notice that this equation is nothing but the Fourier transform of the 
BKP equation which is usually written in the momentum space.
Since the 3-quark odderon operator Eq.~(\ref{sy}) can be represented 
as a linear combination of the 3-point Green's functions, the equivalence
between our result (\ref{prot}) and the BKP equation is essentially 
established. The apparent difference appeared because our operator
partly contains the information of the impact factor of the projectile.

However, there is a caveat when we write 
(\ref{eqf3}). In fact, 
since the Green function $f_\tau(\x,\y,\z)$ is not gauge invariant, 
if one applied the original JIMWLK equation to this operator,
one would obtain a result which is different from (\ref{eqf3}) 
and is even ill-defined due to infra-red divergences.
Instead of doing this, we 
have derived (\ref{eqf3}) from the simplified version of the JIMWLK 
equation which is free of any infra-red divergences and 
is justified for gauge invariant operators. 
This means that we can use the simplified JIMWLK 
equation to gauge {\it variant} operators as far as we finally 
consider gauge invariant quantities. 
In other words, the use of the simplified
JIMWLK equation for the Green function corresponds to a kind of 
regularization.

\vspace{-2mm}


\section*{Acknowledgments}

\vspace{-3mm}
The author is grateful to Yoshitaka Hatta, Edmond Iancu and 
Larry McLerran, with whom the results presented in this talk 
were obtained.\cite{Odderon}

\end{document}